\documentclass[runningheads]{llncs}
\usepackage{graphicx}
\usepackage{tabularx}
\usepackage{xcolor}
\usepackage{multirow}
\usepackage{array, booktabs}
\usepackage{tikz,pgfplots}
\usepackage{xcolor}
\usepackage{subcaption} 
\usepackage[ruled,vlined]{algorithm2e}
\usepackage{amsmath, amssymb}
\usepackage{mdframed}

\begin{document}
\title{Using Social Choice Theory to Finalize Architectural Decisions}
\titlerunning{Non-manipulable Architectural Decisions}
%
%
\author{Shipra Sharma \and
Balwinder Sodhi}
%
\authorrunning{S. Sharma et al.}
%
\institute{Indian Institute of Technology Ropar, India \\
\email{\{shipra.sharma, sodhi\}@iitrpr.ac.in}}
\maketitle              
\begin{abstract}
Unbiased and objective architectural design decisions are crucial for the success of a software development project. Stakeholder inputs play an important role in arriving at such design decisions. However, the stakeholders may act in a biased manner in order to ensure that their requirements and concerns are addressed in a specific way in the software. Most of the existing methods of architectural decision-making do not adequately account for such biased behavior of stakeholders.

We view the software architecture design as a \textit{collective decision making} (CDM) problem in social choice theory, and introduce the central ideas of mechanism design to our field. 
Our contributions are twofold: \textbf{i)} Using the impossibility results from social choice theory, we show that a rational stakeholder can game, to her advantage, almost every known method of software architecture decision making. 
\textbf{ii)} We also show that if the architect is willing to bear the
    an extra cost of giving suitable incentives to stakeholders, then architectural decision making can be protected from strategic (biased) manipulations. For achieving this objective, we introduce the Vickrey-Clarke-Groves (VCG) mechanism from social choice theory to the software architecture domain. 

We illustrate our contributions by comparing and using the examples drawn from the well-known CBAM method of architectural decision making.

\end{abstract}
\section{Introduction}
\label{sec:introduction}
``Architectures are created solely to meet stakeholder needs'' - \cite{rozanski2011software}

Designing architecture is one of the most decisive phases of the software development life cycle. The architecture of a software relates the goals of stakeholders to different aspects of the requirements and then the well-defined requirements to the actual code of the project. This ensures traceability between the software's motivation, objectives, and the final product. The traceability is enabled via software architecture and its related artifacts. 

Architectural design decisions describe the rationale that led to the development of software architecture. Such design decisions were earlier considered as one of the architectural artifacts\cite{van2006design}. However, in recent years a paradigm shift in project development methodologies has made researchers and practitioners to consider the architectural design decisions as software architecture itself \cite{van2016decision}. 

Every architectural decision, and in turn the entire software architecture, should be able to address the validated expectations of the stakeholders and provide a practical solution for it. Therefore, arriving at the architectural decisions is often considered difficult and is fraught with human biases. The human bias can originate from the architect or from stakeholders or both \cite{van2016decision} \cite{tang2017human}. Also, the existing techniques for arriving at software architecture decisions do not have a formal mechanism to handle stakeholders input bias \cite{lopes2017architectural}. To avoid such human bias and to choose an architecture that is fair to all stakeholders we propose a fair model of selecting an architecture.

Various frameworks such as CBAM \cite{kazman2001quantifying} (which is based on ATAM \cite{kazman1999experience}), \cite{letier2014uncertainty}, etc. have been developed for choosing the best candidate architecture from the perspective of modern economic theory. However, a significant assumption taken by these frameworks is the truthfulness of stakeholder inputs, even when it is against the stakeholder's personal benefit. This is in stark contrast to the well-accepted assumption in modern microeconomics that all agents participating in any economic activity are selfish and rational, and are only concerned with their personal benefits \cite{AGT}. Although CBAM-2 \cite{moore2003quantifying} highlights the stakeholder's tendency to manipulate the architectural decisions, it fails to offer a formal mechanism to handle human bias. CBAM-2 merely provides a generic discussion-based approach to handle such situations and they suggest to altogether replace the voting based decision making.

Therefore, we address the following two questions in this paper:
    
{\it Question-1. How do the ``economic frameworks" for choosing a ``most suited" software architecture perform when faced with the possibility of selfish stakeholders, who have the natural objective of influencing the decision processes to their (personal or group) advantage?}
    
{\it Question-2. How can an organization set up incentives to ensure truthful reporting and prevent collusion by various stakeholders while designing software architecture?}
    
To the best of our knowledge, the above questions have not been studied adequately in the software architecture decision making literature. There exists no mechanism which ensures that the stakeholders reveal their true concerns and priorities, especially in relation or response to the other stakeholders' inputs. Using the tools of game theory and mechanism design developed in the field of microeconomics, in this paper, we initiate the first systematic and detailed study of these questions.
    
The rest of the paper is presented as follows.  In  \S \ref{sec:motive} we present our objective by explaining the concept of stakeholder rationality assumption and prove it with the help of a well-known example. \S \ref{sec:socialchoicetheory} explains the social choice theory in the context of architectural design decisions. In order to address the Question-1 that we highlighted in the preceding paragraph, in \S \ref{sec:impossibility} we give a rigorous proof of our claim that all of the architectural design decisions are manipulable. Further, to address the Question-2, we present in \S \ref{sec:VCG} a non-manipulable algorithm that ensures unbiased decisions during the architecture design. \S \ref{sec:relwork} discusses other economic models and frameworks in software architecture domains, and finally, \S \ref{sec:Conclusion} concludes our work and explains our future work.

\section{Motivation and Objective}\label{sec:motive}
Different stakeholders in a software development project usually have different concerns or goals. However, such concerns may overlap each other, or be confirming to each other, or even may conflict. Typically, the benefit of a particular architectural decision to a stakeholder is expected to be aligned with the business, technological or quality goals championed by that stakeholder.

Current methodologies for software architecture design inherently assume that {\it every} stakeholder supplies to the architect an {\it accurate picture} of their inputs and preferences among the various competing architectural designs. In practice, this assumption is usually untenable \cite{mas1995microeconomictheory} as it stipulates that each stakeholder will be truthful about one's inputs, even if it leads to their personal loss. In reality, stakeholders are rational ``players" and they are capable of influencing the architectural decisions for their own benefits. Further, these ``players'', instead of manipulating the game alone, may also act in unison to have an even greater effect on the architectural decisions. We show how such tactics may work in \S \ref{sec:cbam-example} by considering the example of a well-known architecture evaluation method. Before that, we explain the rationality assumption.
\subsection{The stakeholder rationality assumption} 

It is common knowledge and experience that each stakeholder acts rationally and in her own best interest. An important consequence of such rational (or, rather ``selfish'') behavior of the stakeholder is that she may strategically report an inaccurate or suitably modified picture of her true preferences. She may do it to ensure that the overall architectural decisions are more favorable to her own goals. 
    
An even more unfortunate consequence of stakeholder rationality may be {\it collusion}: a group of stakeholders decides together on reporting manipulated preferences so that the architect can be induced to select a particular software architecture design of their liking. 
  
\subsection{Effects of stakeholder rationality on CBAM method}\label{sec:cbam-example}
We now illustrate the impact ``stakeholder rationality assumption" can have on the dynamics and outcome
of a given architecture selection procedure. For this purpose, we take an example with the well-known CBAM methodology \cite{kazman2001quantifying} 
for most suited architecture selection. 

\begin{mdframed}\small
CBAM in a nutshell (Adapted from \cite{kazman2001quantifying})
\tiny
  \begin{enumerate}
    \item Stakeholders report their benefits for every architecture strategy to the architect. 
          We use $r_1, r_2, \ldots, r_n$ to denote the reported benefits. In particular, for each $1 \leq i \leq n$, $r_i(ASj)$ denotes the benefit reported for architecture strategy $ASj$ by stakeholder $s_i$. The benefits are in the range $[-100, 100]$.
          
    \item Costs of various architecture strategies are estimated. Let $c_j$ be the mean cost estimated for
          architecture strategy $ASj$. Usually, there will also be an uncertainty interval around $c_j$.
          
    \item Desirability scores for architecture strategies are computed. The desirability score $d_j$ of $ASj$ is
          equal to average reported benefit divided by mean cost:
                $$d_j = \frac{\sum_{i=1}^{n} r_i(ASj)}{n \cdot c_j}$$
    \item Architecture strategies with high desirability scores are compared based on uncertainty in their costs and benefits.
           
    \item One or more architecture strategies with high desirability score and low uncertainty are selected.
        
    \end{enumerate}
    
    Note: The above outline of the CBAM method is sufficient for the purposes of this paper. However, we would like to note that there is a detailed process attached to each of the above basic steps. 
 \end{mdframed} 
For ease of understanding, we have used the same terminology as used in CBAM for describing the following example. Suppose we have a scenario where the architect must choose an architecture for software such that certain quality attributes (QAs) are satisfied to the desired level. Further, suppose that a method such as CBAM is used to help in making such a selection. To keep the scenario simple, let's say there is only one QA under consideration: {\it battery lifetime}, and assume that there are only three architecture strategies $AS1$, $AS2$, and $AS3$ affecting this QA, and three 
stakeholders $s_1$, $s_2$, and $s_3$ in the decision-making process. The architect has decided to employ
CBAM methodology and will be selecting exactly one architecture strategy out of $AS1$, $AS2$, and $AS3$.

As per CBAM methodology, the architect first asks each stakeholder to report the contribution scores between
$-1$ and $1$ for each candidate architecture strategy. The benefits of a candidate architecture strategy for 
a particular stakeholder is obtained by multiplying contribution score by $100$. Suppose the actual benefits derived are as given in Table-\ref{table-cbam1}.
\begin{table}[h]
\begin{center}
\renewcommand{\arraystretch}{2}
\begin{tabular}{| m{6.5cm}| m{1.5cm} | m{1.5cm} | m{1.5cm} |}
\hline
   & Strategy AS1 & Strategy AS2 & Strategy AS3 \\
\hline
Actual Benefits of stakeholder, $s_1$ & 80 & 70 & 65 \\
\hline
Actual Benefits of stakeholder, $s_2$ & -90 & 50 & 60 \\
\hline
Actual Benefits of stakeholder, $s_3$ & -50 & 50 & 62 \\
\hline
Average Actual Benefit, B & -20 & 56.67 &  62.33 \\
\hline
Anticipated Cost, C & 80 & 95 & 90 \\
\hline
$Desirability = B\div C$ & -0.25 & 0.6 & 0.69  \\
\hline
\end{tabular}
\caption{CBAM based on actual stakeholder benefits for architecture strategies}
\label{table-cbam1}
\end{center}
\end{table}

The average benefit for each architecture strategy is also depicted in Table-\ref{table-cbam1}. The last two rows of this table contain the cost of each architecture strategy, as well as its desirability score as per CBAM methodology.

From the data shown in Table-\ref{table-cbam1}, one can, therefore, conclude that, if all stakeholders had reported their true and correct benefit values to the architect, $AS3$ would have been selected as per the CBAM method. This is because the strategy $AS3$ has the highest desirability. This can be pictorially seen from Fig-\ref{fig:Actual_CBAM_plot}, where cost is on $X$-axis, the benefit is on $Y$-axis, and cost-benefit evaluation of each stakeholder for each architecture strategy is represented by a point.

\begin{figure}[t]
\begin{minipage}[b]{0.35\textwidth}
\begin{tikzpicture}[scale=0.70]
\begin{axis}[legend pos=south west, 
    xmin=0,
    xmax=100,
    ymin=-100,
    ymax=100, %
    xlabel={Cost},
    ylabel={Benefit},
scatter/classes={%
    b={mark=triangle*,red},
    c={mark=square*,red},
    d={mark=o,draw=red,fill=red}
    }]
\addplot[scatter,only marks,%
    scatter src=explicit symbolic]%
table[meta=label] {
x y label
80 80 b
80 -90  b
80 -50 b
95 70 c
95 50 c
95 50 c
90 65 d
90 60 d
90 62 d
    };
    \legend{AS1, AS2, AS3}
\end{axis}
\end{tikzpicture}
\caption{Without Manipulation}
\label{fig:Actual_CBAM_plot}
\end{minipage}
\hspace{2cm}
  \begin{minipage}[b]{0.35\textwidth}
  \begin{tikzpicture}[scale=0.70]
\begin{axis} [legend pos=south west, 
    xmin=0,
    xmax=100,
    ymin=-100,
    ymax=100,
    xlabel={Cost},
    ylabel={Benefit},
scatter/classes={%
    b1={mark=triangle*,blue},
    c1={mark=square*,blue},
    d1={mark=o,draw=blue,fill=blue}
    }]
\addplot[scatter,only marks,%
    scatter src=explicit symbolic]%
table[meta=label] {
x y label
80 80 b1
80 -90  b1
80 -50 b1
95 50 c1
95 50 c1
95 50 c1
90 -10 d1
90 60 d1
90 62 d1
    };
    \legend{AS1, AS2, AS3}
\end{axis}
\end{tikzpicture}
\caption{After Manipulation}
\label{fig:manipulated_cbam_plot}
\end{minipage}
\end{figure}

Now, let us change the scenario to selfish stakeholders, each working towards their own advantage. Suppose stakeholder
$s_1$ somehow gets to know that her first preference $AS1$ will never get selected, as it is scored very
low by the remaining stakeholders. Suppose the next best outcome for $s_1$ is the selection of $AS2$, since she
prefers it over $AS3$. However, the CBAM method with truthful reporting will select her least preferred alternative
$AS3$.

To tip the decision process in favour of $AS2$ over $AS3$, $s_1$ will strategize by falsely reporting a much higher benefit for $AS2$ and a much lower benefit for $AS3$. Suppose $s_1$ reports the benefit of $AS1$ at its actual value $30$, but falsely reports a higher benefit of $50$ for $AS2$ (we are assuming that $s_1$ is very lucky in choice of her false values) and a lower benefit of $-10$ for $AS3$. Further, suppose that the stakeholders $s_2$ and $s_3$ chose to give non-manipulated inputs. The outcome of CBAM method is now given in Table-\ref{table-cbam2}. This situation is pictorially represented in Fig-\ref{fig:manipulated_cbam_plot}.

\begin{table}[h]
\begin{center}
\renewcommand{\arraystretch}{2}
\begin{tabular}{| m{6cm} | m{1.9cm} | m{1.9cm} | m{1.9cm} |}
\hline
   & Strategy AS1 & Strategy AS2 & Strategy AS3 \\
\hline
Reported Benefits of stakeholder, $s_1$ & 80 & \textbf{50} & \textbf{-10} \\
\hline
Reported Benefits of stakeholder, $s_2$ & -90 & 50 & 60 \\
\hline
Reported Benefits of stakeholder, $s_3$ & -50 & 50 & 62 \\
\hline
Average Reported Benefit, B & -20 &  \textbf{50} &  \textbf{37.33} \\
\hline
Anticipated Cost, C & 80 & 95 & 90 \\
\hline
$Desirability = B \div C$ & -0.25 &  \textbf{0.53} &  \textbf{0.41} \\
\hline
\end{tabular}
\caption{CBAM based on reported stakeholder benefits for architecture strategies. Wherever the reported values
are different from actual values, they have been shown in bold font.}
\label{table-cbam2}
\end{center}
\end{table}

Clearly, architecture strategy $AS2$ will be selected instead of $AS3$ now, as it has the maximum desirability
score as well as due to the fact that the uncertainty in the desirability score is zero (all three stakeholders
report the same value of $50$ for benefit from strategy $AS2$). Thus, by strategically misreporting her benefits based on available information about stakeholder choices, stakeholder $s_1$ was able to ensure $AS2$ gets selected instead of $AS3$. Since she prefers $AS2$ more than $AS3$,
she has been able to influence the CBAM method in favor of her preferences.

To summarize, in the event of truthful reporting, $AS3$ would have been the option with maximum desirability score
and minimum risk or uncertainty. However, strategically biased reporting by stakeholder $s_1$, leads to $AS2$ taking the
place of $AS3$ as a preferred option in terms of costs, benefits, and uncertainty. We would like to note that the dynamics of CBAM method will be much more complicated when all stakeholders have access to various amounts of insider information about judgments of other stakeholders, and hence will misreport their benefits, by strategizing based on available information, either in groups or individually.

Given the existing gap in the architectural design decision-making process, we propose a decision-making method using social choice theory. Our approach can be used to find the optimal architecture in a given scenario. We propose an objective voting based mechanism which can avoid manipulation of the architecture selection process. We believe that a discussion-based method for the purpose is often very time consuming and subjective. Also, discussions are difficult to converge in complex scenarios.

\section{Software architecture design in the context of social choice theory}\label{sec:socialchoicetheory}

The main thesis of this paper is that the software architecture design process can be naturally modeled as a \textit{collective decision making (CDM)} problem in social choice theory. In a social choice theory framework, a CDM problem can be described as follows.

We have a set $S'=\{p_1, p_2, \ldots, p_n\}$ of $n$ participants and a finite set $X'=\{x_1, x_2, \ldots, x_m\}$ of $m$ alternatives or outcomes. Each participant $p_i$ has a particular preference order or valuation function $v_i: X' \rightarrow \mathbb{R}$ over the set $X'$ of all alternatives. The objective is to collate all individual preferences or valuations to arrive at a common preference or valuation over the set of all alternatives. Further, the choice of a common preference or valuation over the alternative set $X'$ must be appropriate from the perspective of overall ``social welfare''.

We now compare the above situation with a typical software architecture design process. An architect has to design a particular software architecture which rationally collates the viewpoints and preferences of various stakeholders involved with the software in varying capacities. Examples of stakeholders are acquirers, developers, maintainers, testers, end users, etc. \cite{rozanski2011software}. 

\subsection{CDM modeling}
\label{sec:cdm-mapping}
The mapping to social choice theory is as follows:

\begin{enumerate}
    \item The set of possible software architectures (or architecture strategies) form the set $X=\{a_1, a_2, \ldots, a_m\}$ of alternatives.
    \item The set of different stakeholders form the set $S=\{s_1, s_2, \ldots, s_n\}$ of participants.
    \item Each stakeholder $s_i$ has some benefit (utility) $u_i(x) \in \mathbb{R}$ for every candidate architectural strategy $x \in X$. Alternatively, in case of preferences, each stakeholder $s_i$ has a linear ordering $o_i$ of the set $X$. If $x$ occurs before $x'$ in ordering $o_i$, it means that stakeholder $s_i$ prefers architecture strategy $x'$ over $x$.
    \item The goal of the software architect is to choose an architecture strategy $x^* \in X$ which maximizes a suitable notion of social welfare i.e., an architecture which rationally collates the various preferences received by the architect from the stakeholders. 
\end{enumerate}

\section{Impossibility of designing a non-manipulable software architecture selection method for selfish stakeholders}\label{sec:impossibility}

In \S \ref{sec:cbam-example}, we illustrated by means of an example that the outcome of the well-known CBAM method for software architecture selection can be manipulated by selfish stakeholder(s) in their favor. In light of this example, we now consider the following natural question:

{\it Is there a software architecture selection method which cannot be strategically manipulated by selfish stakeholder(s) in their favor?}

In this section, we answer the above question in the negative i.e., we show that {\it every stakeholder-input based software architecture selection method can be gamed}. This has two consequences:

\begin{enumerate}
    \item There does not exist any CBAM-like method which is non-manipulable, i.e., a method which is immune to stakeholder manipulation as described in  \ref{sec:cbam-example}.
    
    \item The existence of the example discussed in \S \ref{sec:cbam-example} is not particular to the CBAM method; similar examples can be constructed for any other software architecture selection method which selects an optimal architecture based on stakeholders' inputs.
\end{enumerate}

The rest of this section is organized as follows.
We start by providing some definitions and setting up relevant notation. In \S \ref{selection1}, we provide an abstract definition of a software architecture selection method, and we also give a rigorous definition of a non-manipulable software architecture selection method. (Such methods are called {\it dominant strategy incentive-compatible} in
social choice theory literature \cite{AGT}\cite{garg2008foundationsI}\cite{mas1995microeconomictheory}.)

Finally, in \S \ref{impossibility}, we show that no non-trivial non-manipulable software architecture
selection method exists.

\subsection{Definitions}
\label{selection1}

\begin{definition} 
{\bf Software architecture selection method.}
A software architecture selection method $\mathcal{M}$ with stakeholder inputs consists of the following sequence of steps:

\begin{enumerate}
    \item The architect requests each stakeholder $s_i$ to provide her \textbf{actual benefit} $\mathbf{u_i} : X \rightarrow \mathbb{R}$.
    
    \item \textbf{Reported benefit} by each stakeholder is $\mathbf{r_i}: X \rightarrow \mathbb{R}$. $r_i$ takes into consideration the fact that a selfish stakeholder may misreport the actual or true benefit.
   
    \item By applying her decision algorithm $\mathcal{A}$ with the reported benefits $\{r_i ~| ~1 \leq i \leq n\}$ as input, the architect selects a single architecture $a_{i^*}$ from the set $X$ of candidate architectures.
\end{enumerate}
\end{definition}

The key difference among the various software architecture selection methods lies in the design of the algorithm $\mathcal{A}$. For a CDM based on social choice theory, the algorithm $\mathcal{A}$ is called the {\it social choice function} \cite{AGT,garg2008foundationsI}.

\begin{definition}
{\bf Non-manipulable software architecture selection.}
A software architecture selection method $\mathcal{M}$ is called {\it non-manipulable} if and only if it is in the best interest of each stakeholder to report her true benefit to the architect i.e., for all $1 \leq i \leq n$ and for all possible reported benefit functions $r_1, r_2, \ldots, r_n$, we have that:

$$u_i(\mathcal{A}(r_1, r_2, \ldots, u_i, \ldots, r_n)) \geq u_i(\mathcal{A}(r_1, r_2, \ldots, r_i, \ldots, r_n))$$ 
\end{definition}

Note that the above condition ensures truthfulness on the part of each rational stakeholder, irrespective of her awareness about the strategies as well as the actions of other stakeholders. 

\subsection{Non-existence of a non-manipulable software architecture selection method}
\label{impossibility}

The non-existence of non-manipulable software architecture selection methods follows from the Gibbard-Satterthwaite theorem \cite{AGT}\cite{garg2008foundationsI} in social choice theory. For the scenario of software architecture selection, we adapt and state the theorem as follows:

\begin{theorem}
\label{gs} 
Suppose there are at least three different candidate architectures in set $X$. Further, suppose the social choice algorithm $\mathcal{A}$ used by the architect is non-manipulable. 

Then $\mathcal{A}$ is a dictatorship, i.e., there exists an index $i$, $1 \leq i \leq n$, such that the architect always chooses an architecture which maximizes the reported benefit of stakeholder $s_i$ and completely ignores the reported benefits provided by the remaining stakeholders $s_j, j \neq i$.
\end{theorem}

We now illustrate Theorem \ref{gs} with two examples. 

{\it Example 1. Dictatorial CBAM.} Consider a dictatorial version of CBAM, where the architect takes the benefit reported by stakeholder $s_2$ as the average reported benefit, and then selects the architecture with maximum desirability. Thus, the benefits reported by $s_1$ and $s_3$ are discarded by the architect and not used in the decision-making process.

In this selection method, there is no incentive for $s_1$ and $s_3$ to misreport their benefits, as they are anyway not
part of the decision process. Thus, this method cannot be manipulated by $s_1$ and $s_3$. However, $s_2$ can still manipulate the method's outcome, as shown by Table-\ref{dict-cbam}. If $s_2$ had reported her true benefit values, strategy $AS1$ with the desirability of $1.25$ would have been selected. (This is because, though $s_2$ has more benefit from $AS2$ than $AS1$, the cost of $AS1$ is much lower than that of $AS2$.) However, $s_2$ prefers $AS2$ more than $AS1$, and hence by misreporting a higher benefit of $100$ for $AS2$ and a lower benefit of $30$ for $AS1$, she ensures that $AS2$ has the highest desirability and gets selected instead of $AS1$.
\begin{table}[h]

\begin{center}
\renewcommand{\arraystretch}{2}
\begin{tabular}{| m{5.9cm} | m{2cm} | m{2cm} | m{2cm} |}
\hline
   & Strategy AS1 & Strategy AS2 & Strategy AS3 \\
\hline
Actual Benefits of stakeholder, $s_2$ & 50 & 80 & -10 \\
\hline
Reported Benefits of stakeholder, $s_2$ & {\bf 30} & {\bf 100} & -10 \\
\hline
Anticipated Cost, C & 40 & 100 & 50 \\
\hline
Actual Desirability   $=$   Average Actual Benefit $\div$ C & 1.25 & 0.8 & -0.2 \\
\hline
Calculated Desirability $=$ Average Reported Benefit $\div$ C & {\bf 0.75} & {\bf 1} & -0.2 \\
\hline
\end{tabular}
\caption{Manipulating dictatorial CBAM where the benefits reported by $s_2$ are taken as average actual benefits}
\label{dict-cbam}
\end{center}
\end{table}

{\it Example 2. Plain Dictatorship.} Now consider the same example as above, with the following change in selection algorithm $\mathcal{A}$.
The benefits reported by $s_1$ and $s_3$ are discarded. However, now the architecture selected is the one with the maximum reported benefit for $s_2$.

Clearly, $s_1$ and $s_3$ have no incentive to misreport, as their feedback does not count. But now, $s_2$ also has no incentive to misreport her benefit values, as the strategy with maximum reported benefit value for $s_2$ always gets selected. Thus, this method is non-manipulable.

Thus, Theorem \ref{gs} puts serious limitations on CDM and shows that only software architecture selection methods similar to those in Example 2 
above are non-manipulable. Any deviation from plain dictatorship (such as for example 1 above, or the example in \S \ref{sec:cbam-example}), leads to scenarios where stakeholder(s) can unfairly influence the outcome in their favor.

\section{Vickrey-Clarke-Groves (VCG) mechanism for software architecture selection}\label{sec:VCG}

We now propose a novel non-manipulable software architecture selection method $\mathcal{M}$, using the well-known VCG mechanism \cite{AGT},\cite{garg2008foundationsI},\cite{mas1995microeconomictheory}. The non-manipulability of the proposed approach is ensured by making payments to individual stakeholders; the payments are devised in such a way that a selfish stakeholder will always report her true benefit values. 

\begin{figure}[t]
    \centering
    \includegraphics[scale=0.5]{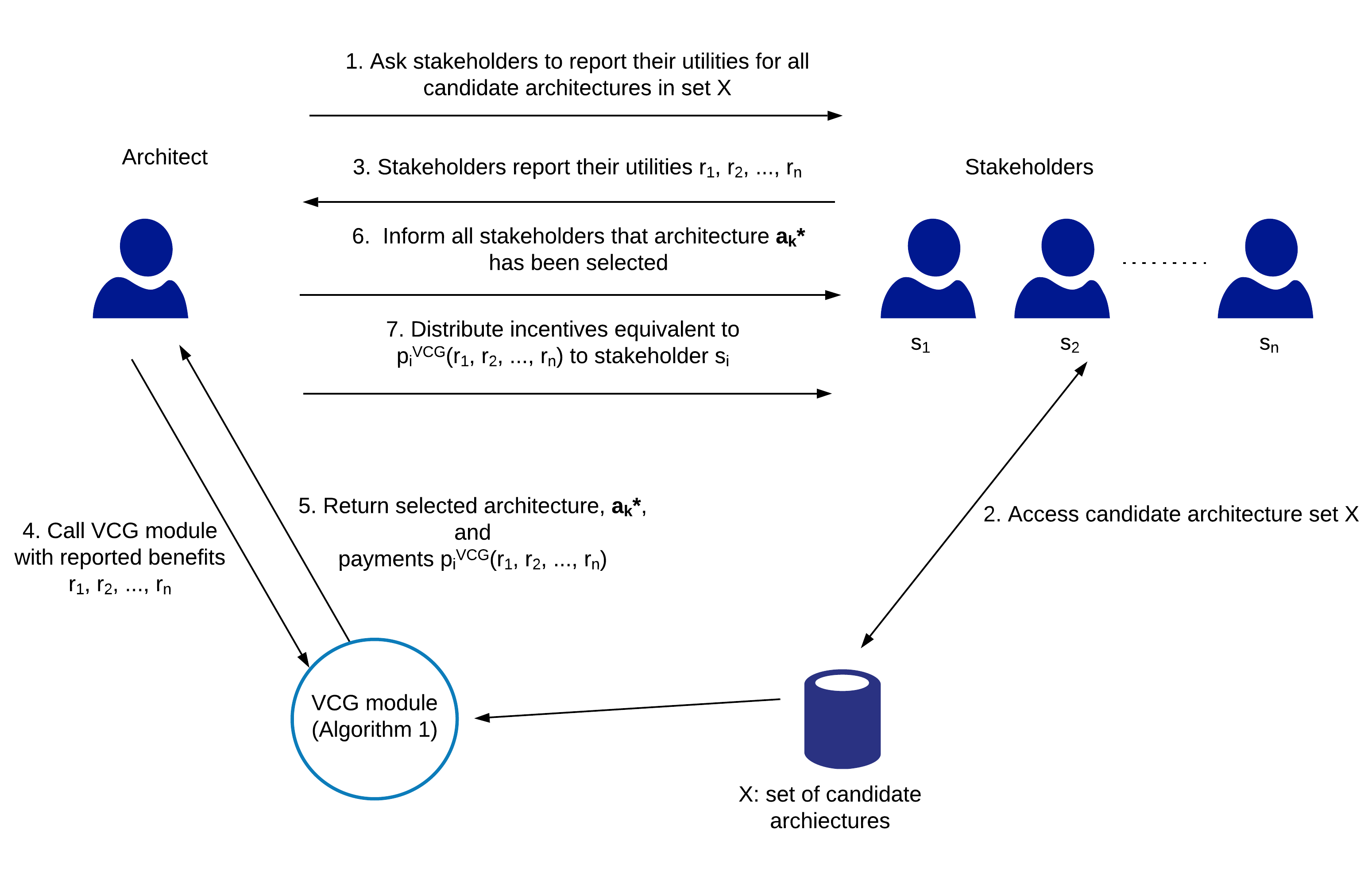}
    \caption{Overview of our non-manipulable architecture selection technique $\mathcal{M}$}
    \label{fig:vcg-overview}
\end{figure}

Figure-\ref{fig:vcg-overview} depicts the proposed non-manipulable approach to selecting architecture using VCG. The numbers on the arrows in the figure depict the sequence of actions for selection. Our approach will always select the architecture with a maximum total reported benefit.  Here, $u_i: X \rightarrow \mathbb{R}, \forall i \in [1, n]$,  denotes the actual benefits and $r_i: X \rightarrow \mathbb{R}$ denotes the reported benefits of stakeholder $s_i$.   

The VCG module depicted in Figure-\ref{fig:vcg-overview} is discussed in Algorithm \ref{Algo:vcg4sa}. For our algorithm (Algorithm \ref{Algo:vcg4sa}) we use Clarke pivot rule \footnote{Note that the payment calculation formula consists of two terms: the first term $T^+_i(r_1, r_2, \ldots, r_n)$ depends on the benefits reported by stakeholder $s_i$; however, the second term $T^-_i(r_1, r_2, \ldots, r_n)$ is due to Clarke pivot rule. This second term is independent of the benefits $r_i$ reported by stakeholder $s_i$. All payments in Clarke pivot rule are negative or zero. Hence, the architect can correspondingly decide the incentives. The reader is referred to \cite{AGT}, \cite{garg2008foundationsI} for further details on Clarke pivot rule.} with VCG. 

\begin{algorithm}[t]
 \caption{{\bf Steps of $\mathcal{VCG}$ module for software architecture selection} \label{Algo:vcg4sa}}
\KwIn{ 1) Reported benefits: $(r_1, r_2, \ldots, r_n)$\\
~~~~~~~~~~~2) $X$: Set of candidate architectures} 

\vspace{0.1cm}

\KwOut{Selected architecture $\mathcal{VCG}(r_1, r_2, \ldots, r_n)$ and payments $p_i^{\mathcal{VCG}}(r_1, r_2, \ldots, r_n)$ for $1 \leq i \leq n$. }

\vspace{0.1cm}

\nl  Calculate the total reported benefit $TRB_k(r_1, r_2, \ldots, r_n)= \sum_{i=1}^{n} r_i(a_k)$ for each candidate architecture $a_k$.
   
\nl Let $a_{k^*}$ be the candidate architecture with the maximum $TRB_k(r_1, r_2, \ldots, r_n)$. If there are more than one such architectures, choose one of them arbitrarily.

\nl For each stakeholder $s_i$, calculate:
    
         $$T^+_i(r_1, r_2, \ldots, r_n) = \sum_{j \neq i} r_j(a_{k^*})$$ $$T^-_i(r_1, r_2, \ldots, r_n) = max_{k \in [1,m]} \Big( \sum_{j \neq i} r_j(a_{k}) \Big)$$

\nl Calculate payment $p_i^{\mathcal{VCG}}(r_1, r_2, \ldots, r_n) = T^+_i(r_1, r_2, \ldots, r_n) - T^-_i(r_1, r_2, \ldots, r_n)$ to stakeholder $s_i$.

\nl  Set $\mathcal{VCG}(r_1, r_2, \ldots, r_n)=a_{k^*}  $.
\end{algorithm}

As shown in Figure-\ref{fig:vcg-overview}, after the VCG module is executed, the architect informs the stakeholders about the selected candidate architecture $\mathcal{VCG}(r_1, r_2, \ldots, r_n)$ and further distributes net incentives equivalent to $p_i^{\mathcal{VCG}}(r_1, r_2, \ldots, r_n)$ to each stakeholder $s_i$. Each stakeholder derives benefits due to two reasons: (i) an actual benefit of $u_i( \mathcal{VCG}(r_1, r_2, \ldots, r_n)  )$ due to the selection of the particular architecture by VCG module, and
(ii) incentives to the value of $p_i^{\mathcal{VCG}}(r_1, r_2, \ldots, r_n)$ from the architect. Thus, the
{\it net benefit} $NB_i^{\mathcal{VCG}}(r_1, r_2, \ldots, r_n)$ of stakeholder $s_i$ under $\mathcal{VCG}$ mechanism is equal to $u_i(a_{k^*}) + p_i^{\mathcal{VCG}}(r_1, r_2, \ldots, r_n)$. As the stakeholder is rational,
her objective is now to maximize her net benefit. Thus, the following social choice theory theorem (Theorem-\ref{theorem:t2}), which is stated from \cite{AGT}\cite{garg2008foundationsI}, applies in this context.

\begin{theorem}\label{theorem:t2}
    If the objective of each selfish stakeholder is to maximize her {\it net benefit}, then $\mathcal{VCG}$ is non-manipulable.
    \label{vcg-nm}
\end{theorem}

Here, the ``net benefit'' represents the stakeholder benefit after incorporating her incentive payments. Throughout this paper, we have used the term ``benefits'' to indicate just the direct benefits expected by a stakeholder when a specific architecture alternative gets selected. 

It is pertinent here to note that if we replace ``net benefit'' with ``benefit'' in Theorem \ref{theorem:t2}, then the mechanism becomes manipulable. 

\subsection{Examples of software architecture selection using $\mathcal{VCG}$ algorithm}

We now illustrate the outcome of architecture selection by $\mathcal{VCG}$ on the examples considered earlier in this paper.

\subsubsection{CBAM Example from \S \ref{sec:cbam-example}.}

We now implement $\mathcal{VCG}$ mechanism on the CBAM example given in \S \ref{sec:cbam-example}.
    
Let $u_1$, $u_2$, and $u_3$ be the actual benefits of stakeholders $s_1$, $s_2$, and $s_3$ (see first three rows of Table-\ref{table-cbam1}). Further, let $r_1$, $r_2$, and $r_3$ denote the reported benefits of stakeholders $s_1$, $s_2$, and $s_3$ (see first three
rows of Table-\ref{table-cbam2}). Table-\ref{table-vcg-cbam1} describes the computation of payments $p_i^{\mathcal{VCG}}(r_1, r_2, r_3)$ and net benefits $NB_i^{\mathcal{VCG}}(r_1, r_2, r_3)$.
 
\begin{table}[t] \begin{minipage}{.5\linewidth}
    \begin{center}
        
        \begin{tabular}{|c|m{1cm}|m{1cm}|m{1cm}|}
            \hline
            $i$ & $s_1$  \newline ($i=1$) & $s_2$ \newline ($i=2$) & $s_3$ \newline ($i=3$) \\ \hline
            $T_i^+(r_1, r_2, r_3)$ & 100 & 100 & 100 \\ \hline
            $T_i^-(r_1, r_2, r_3)$ & 122 & 100 & 100 \\ \hline
            $p_i^{\mathcal{VCG}}(r_1, r_2, r_3)$ & -22 & 0 & 0 \\ \hline
            $u_i(\mathcal{VCG}(r_1, r_2, r_3))$ & 50 & 50 & 50 \\ \hline
            $NB_i^{\mathcal{VCG}}(r_1, r_2, r_3)$ &28 & 50 & 50 \\ \hline
        
        \end{tabular}
        \caption{Computation of payments and net benefits as per $\mathcal{VCG}$ mechanism, based on reported benefits $r_1$, $r_2$, and $r_3$.}
        \label{table-vcg-cbam1}
    \end{center}
    \end{minipage}
    ~
    \begin{minipage}{.5\linewidth}
    \begin{center}
        \begin{tabular}{|c|m{1cm}|m{1cm}|m{1cm}|}
            \hline
            $i$ & $s_1$ \newline ($i=1$) & $s_2$ \newline ($i=2$) & $s_3$ \newline ($i=3$) \\ \hline
            $T_i^+(u_1, r_2, r_3)$ & 122 & 127 & 125 \\ \hline
            $T_i^-(u_1, r_2, r_3)$ & 122 & 127 & 125 \\ \hline
            $p_i^{\mathcal{VCG}}(u_1, r_2, r_3)$ & 0 & 0 & 0 \\ \hline
            $u_i(\mathcal{VCG}(u_1, r_2, r_3))$ & 65 & 60 & 62 \\ \hline
            $NB_i^{\mathcal{VCG}}(u_1, r_2, r_3)$ &65 & 60 & 62 \\ \hline
        \end{tabular}
        \caption{Computation of payments and net benefits as per $\mathcal{VCG}$ mechanism, based on reported benefits $u_1$, $r_2$, and $r_3$.}
        \label{table-vcg-cbam2}
    \end{center}\end{minipage}
\end{table}
    
Now, suppose stakeholder $s_1$ unilaterally decides to report her true benefit values. Thus, the reported benefits are $u_1$, $r_2$, and $r_3$. The outcome of $\mathcal{VCG}$ mechanism is summarized in Table-\ref{table-vcg-cbam2}.

Thus, if stakeholder $s_1$ reports her benefit values truthfully to mechanism $\mathcal{VCG}$, and the other stakeholders do not change their benefit values, the net benefit of $s_1$ increases from $28$ to $65$. Hence, if the architect had implemented an algorithm $\mathcal{VCG}$, it would have been in the best interest of $s_1$ not to misreport her benefit values. Since $\mathcal{VCG}$ algorithm will ensure that each stakeholder speaks the truth, the architecture strategy $AS3$ will be selected by it since it has the maximum total reported (which would now be actual) benefit. 

Note that in this particular example, the architecture strategy selected by $\mathcal{VCG}$ is the same
as that selected by CBAM. However, this is not always the case.

\subsubsection{Dictatorial CBAM example from \S \ref{impossibility}.}

Now consider the dictatorial CBAM method of \S \ref{impossibility}. Further, suppose that the actual benefits $u_2$ and reported benefit $r_2$ of stakeholder $s_2$ are those given in first two rows of Table-\ref{dict-cbam}. To complete the input specification, we assume that the actual benefit vectors of $s_1$ and $s_3$ are $u_1=(55, ~60, ~10)$ and $u_3=(~40, ~50, ~-20)$ respectively. Table-\ref{table:dict-table-vcg-cbam1} describes the actual run of $\mathcal{VCG}$ for reported benefits $u_1$, $r_2$, and $u_3$. 

\begin{table}[b]\begin{minipage}{.5\linewidth}
    \begin{center}
        \begin{tabular}{|c|m{1cm}|m{1cm}|m{1cm}|}
            \hline
            $i$ & $s_1$ \newline ($i=1$) & $s_2$ \newline ($i=2$) & $s_3$ \newline ($i=3$) \\ \hline
            $T_i^+(u_1, r_2, u_3)$ & 150 & 110 & 160 \\ \hline
            $T_i^-(u_1, r_2, u_3)$ & 150 & 110 & 160 \\ \hline
            $p_i^{\mathcal{VCG}}(u_1, r_2, u_3)$ & 0 & 0 & 0 \\ \hline
            $u_i(\mathcal{VCG}(u_1, r_2, u_3))$ & 60 & 80 & 50 \\ \hline
            $NB_i^{\mathcal{VCG}}(u_1, r_2, u_3)$ & 60 & 80 & 50 \\ \hline
        \end{tabular}
        \caption{Computation of payments and net benefits as per $\mathcal{VCG}$ mechanism, based on reported benefits $u_1$, $r_2$, and $u_3$.}
        \label{table:dict-table-vcg-cbam1}
    \end{center}
    \end{minipage}
~
\begin{minipage}{.5\linewidth}
    \begin{center}
        \begin{tabular}{|c|m{1cm}|m{1cm}|m{1cm}|}
            \hline
            $i$ & $s_1$ \newline ($i=1$) & $s_2$ \newline ($i=2$) & $s_3$ \newline ($i=3$) \\ \hline
            $T_i^+(u_1, u_2, u_3)$ & 130 & 110 & 140 \\ \hline
            $T_i^-(u_1, u_2, u_3)$ & 130 & 110 & 140 \\ \hline
            $p_i^{\mathcal{VCG}}(u_1, u_2, u_3)$ & 0 & 0 & 0 \\ \hline
            $u_i(\mathcal{VCG}(u_1, u_2, u_3))$ & 60 & 80 & 50 \\ \hline
            $NB_i^{\mathcal{VCG}}(u_1, u_2, u_3)$ & 60 & 80 & 50 \\ \hline
        \end{tabular}
        \caption{Computation of payments and net benefits as per $\mathcal{VCG}$ mechanism, based on reported benefits $u_1$, $u_2$, and $u_3$.}
        \label{table:dict-table-vcg-cbam2}
    \end{center}
\end{minipage}
\end{table}
Now suppose stakeholder $s_2$ had instead reported her true benefit values $u_2 = (50, ~80, -10)$. Table-\ref{table:dict-table-vcg-cbam2} describes a run of algorithm $\mathcal{VCG}$ with reported benefits $u_1$, $u_2$, and $u_3$. Since speaking the truth keeps the net benefit of $s_2$ the same, there is no incentive for her to lie under mechanism $\mathcal{VCG}$. Further, since all stakeholders are constrained to report true benefits, $\mathcal{VCG}(u_1, u_2, u_3)$ is $AS2$, and this outcome is non-manipulable.

\subsection{Discussion}
\subsubsection{Why a utilitarian objective function?}

In algorithm $\mathcal{VCG}$ above, the architect always picks the architecture with maximum total reported benefit. The sum of all stakeholder benefits for a given architecture is called the utilitarian objective function. A natural question is whether VCG-type mechanisms can be devised which select an architecture based on objective functions other than the utilitarian objective function while ensuring truthfulness by
rational stakeholders.

There is considerable literature in social choice theory and mechanism design on how to deal with this question. 
A theorem due to Roberts (\cite{AGT}), proves that if there is no restriction on the possible benefit functions, then the only possible social choice functions for which the above VCG-type decision algorithm can force stakeholder truthfulness are {\it weighted utilitarian functions}. Essentially, these are social choice functions which always choose an architecture which maximizes the weighted sum of reported benefits of a particular subset of stakeholders. In light of these observations, the objective function chosen is utilitarian.

\subsubsection{How can the architect incentivize the decision process?}

Usually, the form of incentives are specific to an organization and its culture. For example, when all stakeholders are members of a single organization, the payments to be derived from the various stakeholders by algorithm $\mathcal{VCG}$ can be implemented by the administrators using absence or presence of certain organizational incentives such as higher salary, bonus, or shares, elevation or promotion, commendations, extra fund allocation, etc. 

In a situation where some or all aspects of software development are being carried out by outside organizational entities (third party), the use of algorithm $\mathcal{VCG}$ by the architect will encourage truthfulness on part of various competing entities and lead to maximization of total benefits. The payments relative to outside entities stipulated by algorithm $\mathcal{VCG}$ can then be implemented using certain profitable contract clauses, long-term service contracts, recommendation for future projects, etc.

\section{Related Work}\label{sec:relwork}
To the best of our knowledge, there exists no decision-making technique in software architecture which uses mechanism design or social choice theory. Hence, we consider those methodologies related to our work which takes stakeholders' concerns as input along with requirements and other environmental constraints. 

In methods like CBAM \cite{kazman2001quantifying}, \cite{asundi2001using} and \cite{kazman2002making} the authors made the customer decide which stakeholder to be included in the decision-making process, and then these stakeholders give some numeric number or score to the quality attributes, where these number should sum to 100. They made the stakeholder describe why or how a particular quality is essential. To choose between the various architectural strategy they calculate a desirability factor which decreases with cost. CBAM paper lists three important criteria to arrive at an architectural decision: costs, benefits, and uncertainty or risks. All data gathered and used during an architecture decision process is at best only an estimate, and therefore is best represented by associated probability distributions, rather than exact, definite values. The CBAM framework measures the relevance of a candidate architecture by the ratio of benefits to costs. 

In \cite{letier2014uncertainty} the architects in collaborations with project stakeholders define the cost and benefit functions. Each alternative has a utility score defined as the weighted sum of the stakeholders' preferences for each goal.  Hence they measure the relevance of a candidate architecture by the ``net benefit,'' i.e., quantity obtained by subtracting the costs from the benefits. CEDA \cite{nakakawa2009requirements} considers stakeholder's concerns to develop shared understanding aligned towards an organization's goals. Tools like RADAR \cite{busari2017radar} use CBAM or ATAM as the starting point of developing an architecture and then handle particular aspects of architecture design decision like uncertainty using a Pareto optimal approach. Techniques like REARM\cite{martinez2013rearm} accept quantifiable input from stakeholders, but do not provide the details of how is to be done. Further, techniques like \cite{lopes2017architectural} integrate updated versions of CBAM-2 to handle structured analysis of architecture in agile development. \cite{schenkhuizen2016consistent} suggests a workshop based approach to handle concerns of stakeholders and specifically mentions as automating the handling of concerns and conflict as the need of the hour.


\section{Conclusion and Future Work}\label{sec:Conclusion}
Software  architecture comprises of a set of decisions that need to be taken timely and correctly to avoid project failures.  Reaching these decisions involves handling of the requirement of various stakeholders, who may (rightfully) push for their concerns to be realized in a certain way in software architecture. In software architecture research mostly this human aspect of architectural decisions is ignored. 

We provide a new paradigm in software architecture decision-making by objectively handling human biases such as personal benefits, habits, strategically biased inputs, and so on. We identify the presence of rational stakeholders in the architectural decision-making process and propose the usage of social choice theory and mechanism design to handle their concerns. We provide a rigorous framework to ensure truthful information elicitation from rational stakeholders, thus ensuring that human factors do not influence software architecture design.

We are working on developing a tool, based on our proposed and validated VCG methodology to assist the architect in eliciting truthful information from stakeholders and thus make unbiased decisions while designing software architecture. This tool would be used by an architect to resolve the conflicts and concerns of stakeholders and automate the process of fair selection of architectural strategies.

\bibliographystyle{splncs04}
\bibliography{my_ref.bib}
\end{document}